\documentclass[prd,showpacs,twocolumn]{revtex4}

\usepackage{hyperref}
\usepackage{amssymb}
\usepackage{amsmath}

\begin{document}

\title{The derivation of the Wheeler-DeWitt equation of $f(R,L_m)$ gravity in minisuperspace}

\author{Ru-Nan Huang}
 \email{huangrn@shu.edu.cn}
 \affiliation{Department of Physics, Shanghai University, Shanghai, 200444, P.R.China}

\date{\today}

\begin{abstract}
In this paper we present the Wheeler-DeWitt equation of $f(R,L_m)$ gravity in flat FRW universe, which is the first step of the study of quantum $f(R,L_m)$ cosmology. In the minisuperspace spanned by the FRW scale factor $a$ and the Ricci scalar $R$, we explicitly examine and verify the equivalence of the reduced action. We then canonical quantize the $f(R,L_m)$ model and derive the corresponding Wheeler-DeWitt equation.
\end{abstract}


\maketitle


\section{Introduction}

From cosmic observations, such as the observations of type Ia supernovae \cite{supernova Ia}, cosmic microwave background radiation \cite{cmb}, large scale structure \cite{lss} and baryon acoustic oscillations \cite{bao}, we know that our universe is currently accelerating. This acceleration is due to the mysterious presence of dark energy, which accounts for 70\% of the cosmic energy content \cite{dark energy}. $f(R)$ gravity, first proposed in \cite{f(R) first proposed}, where the standard Einstein-Hilbert action is replaced by an arbitrary function of the Ricci scalar $R$, now acts as a promising way to solve the dark energy problem \cite{f(R) now}. Viable models that meet the severe weak field criteria in the solar system regime have been given in \cite{wf}. One can see \cite{f(R) review} for a review of $f(R)$ gravity.

Recently, Nojiri et al. generalized the $f(R)$ gravity by including in the action an explicit coupling between geometry and matter, i.e., they wrote the action as \cite{Nojiri:2004bi}
\begin{equation*}
 S=\int d^4x\sqrt{-g}
 \left\{\frac{1}{\kappa^2}R+\left(\frac{R}{\mu^2}\right)^\alpha L_d\right\}
\end{equation*}
where $L_d$ is the matter lagrangian density. The cosmological implications of such non-minimal coupling were widely investigated in \cite{study1}. It is revealed that the non-minimal coupling will give rise to an extra force on massive particles, which is orthogonal to the four velocities, and consequently cause non-geodesic motions. Harko and Lobo further extended such models to the maximal extension of the Einstein-Hilbert action, i.e., they wrote the action as
\begin{equation*}
 S=\int d^4x\sqrt{-g} f(R,L_m)
\end{equation*}
where $f(R,L_m)$ is an arbitrary function of the Ricci scalar $R$ and the matter lagrangian density $L_m$ \cite{Harko:2010mv}. The properties of $f(R,L_m)$ gravity are extensively studied in \cite{study2}.


Quantum gravity, proposed by DeWitt in his pioneering paper \cite{DeWitt:1967yk} in 1967, is an effort to provide a quantum theory of the gravitational field and to cast light on the study of cosmology \cite{quancos}. In the canonical quantization approach, which starts with a split of spacetime into space and time, the wave function $\Psi(g_{ij})$ is defined on a superspace, which, according to Wheeler, is an infinite-dimensional manifold whose points correspond to metrics on a three-dimensional spatial foliation, and acts as an arena where the development of the geometries of general relativity takes place \cite{Wheelersuper}. Quantum geometrodynamics and loop quantum gravity are the popular examples of canonical quantum gravity.

Our main purpose in this paper is to present the Wheeler-DeWitt equation of $f(R,L_m)$ gravity in the flat FRW universe. The Wheeler-DeWitt equation plays the role of Schr\"odinger equation for the wave function of the universe, and is at the heart of every canonical approach of quantum cosmology. Since the three-geometry of the flat FRW universe is homogeneous and isotropic, there are no inhomogeneous fluctuations of the metric. Thus the infinite number of gravitational degrees of freedom are reduced to just two variables, the FRW scale factor $a$ and the Ricci scalar $R$, and the resulting configuration space is called minisuperspace. Adopting the method proposed by Vilenkin in \cite{Vilenkin:1985md}, we present the Wheeler-DeWitt equation of $f(R,L_m)$ gravity, which serves as the first step of the study of quantum $f(R,L_m)$ cosmology. 


The paper is organized as follows. In Sec.~\ref{sec:fgravity} we give a brief review of $f(R,L_m)$ gravity developed in~\cite{Harko:2010mv}. Sec.~\ref{sec:redaction} are devoted to the derivation of the reduced action of $f(R,L_m)$ gravity in minisuperspace. We first write the action of $f(R,L_m)$ gravity in flat FRW universe via a Lagrangian multiplier and integrate by parts to get a reduced action. We then apply the variation priciple to obtain the reduced field equations. We explicitly verify that these reduced equations are the same as the field equation in Sec.~\ref{sec:fgravity} under flat FRW metric, which vindicates the equivalence of the reduced action and the original one. In Sec.~\ref{sec:wdequation} we canonical quantize the $f(R,L_m)$ gravity model in minisuperspace and obtain the corresponding Wheeler-DeWitt equation. Sec.~\ref{sec:conclusion} contains a discussion and the conclusions. We use the natural unit system $8\pi G=c=\hbar=1$ throughout this paper.


\section{Brief review of $f(R,L_m)$ gravity}\label{sec:fgravity}

The maximal extension of the Einstein-Hilbert action for non-minimal coupling gravity is given by~\cite{Harko:2010mv}
\begin{equation}\label{eqn:oriaction}
 S=\int d^4x\sqrt{-g}f(R,L_m)
\end{equation}
where $f(R,L_m)$ is an arbitrary function of the Ricci scalar $R$ and the matter Lagrangian density $L_m$. The energy-momentum tensor is defined as usual
\begin{equation}\label{eqn:emtensor}
 T_{\mu\nu}=-\frac{2}{\sqrt{-g}}\frac{\delta(\sqrt{-g}L_m)}{\delta g^{\mu\nu}}
\end{equation}
Assuming that $L_m$ depends only on the metric $g_{\mu\nu}$ and not on their derivatives, (\ref{eqn:emtensor}) can be reduced into
\begin{equation}\label{eqn:redemtensor}
 T_{\mu\nu}=g_{\mu\nu}L_m-2\frac{\partial L_m}{\partial g^{\mu\nu}}
\end{equation}

Varying the action (\ref{eqn:oriaction}) with respect to the metric $g^{\mu\nu}$ and using the expression of $T_{\mu\nu}$ given by (\ref{eqn:redemtensor}), one obtains the field equations of the $f(R,L_m)$ model
\begin{multline}\label{eqn:orifieldeq}
 f_R R_{\mu\nu}+(g_{\mu\nu}\square-\nabla_\mu\nabla_\nu)f_R-\frac{1}{2}(f-f_{L_m}L_m)g_{\mu\nu} \\
 =\frac{1}{2}f_{L_m}T_{\mu\nu} 
\end{multline}
where $\square=g^{\mu\nu}\nabla_\mu\nabla_\nu$ and
\begin{equation}
 f_R \equiv \frac{\partial f(R,L_m)}{\partial R}, \quad
 f_{L_m} \equiv \frac{\partial f(R,L_m)}{\partial L_m}
\end{equation}
Choosing $f(R,L_m)=\frac{1}{2}R+L_m$, i.e., the Einstein-Hilbert action, (\ref{eqn:orifieldeq}) can be reduced to the Einstein equation, $R_{\mu\nu}-\frac{1}{2}g_{\mu\nu}R=T_{\mu\nu}$, just as expected.

\smallskip

Using the following mathematical identities~\cite{Misner:1974qy}
\begin{equation}
 (\square\nabla_\mu-\nabla_\mu\square)f_R=R_{\mu\nu}\nabla^\nu f_R
\end{equation}
and
\begin{equation}
 \nabla^\mu G_{\mu\nu}=\nabla^\mu(R_{\mu\nu}-\frac{1}{2}g_{\mu\nu}R)=0
\end{equation}
one can prove that
\begin{equation}\label{eqn:idcovfR}
 \nabla^\mu\left[f_R R_{\mu\nu}
 +(g_{\mu\nu}\square-\nabla_\mu\nabla_\nu)f_R
 -\frac{1}{2}g_{\mu\nu}f\right]=0
\end{equation}
Taking the covariant derivative of (\ref{eqn:orifieldeq}) and using the identity (\ref{eqn:idcovfR}), one obtains the equation for the divergence of the energy-momentum tensor $T_{\mu\nu}$
\begin{equation}
 \nabla^\mu T_{\mu\nu}=2(\nabla^\mu\ln f_{L_m})\frac{\partial L_m}{\partial g^{\mu\nu}}
\end{equation}
It is shown that in modified gravity with non-minimal coupling between matter and geometry, both the matter Lagrangian and the energy-momentum tensor are completely determined by the form of the coupling \cite{Harko:2010zi}. Once the matter Lagrangian density $L_m$ is given, one can choose appropriate forms of $f(R,L_m)$ to obtain conservative models which give $\nabla^\mu T_{\mu\nu}=0$.


\section{Reduced Lagrangian in minisuperspace}\label{sec:redaction}

We now consider the $f(R,L_m)$ gravity in flat FRW universe filled with matter and write the metric as
\begin{equation}\label{eqn:FRW}
 ds^2=-dt^2+a^2(t)d\vec{x}^2
\end{equation}
which is homogeneous and isotropic and thus drasticlly reduced the gravitational degrees of freedom. This spacetime geometry is consistent with the present cosmological observations, such as the cosmic microwave background radiation, which has deviations from perfect smoothness on the order of $10^{-5}$ \cite{cmb}.

The scalar curvature for the flat FRW metric (\ref{eqn:FRW}) is
\begin{equation}\label{eqn:Ricci}
 R=6(\frac{\ddot{a}}{a}+\frac{\dot{a}^2}{a^2})
\end{equation}
Plugging (\ref{eqn:FRW}) and (\ref{eqn:Ricci}) directly into the $f(R,L_m)$ gravity model (\ref{eqn:oriaction}) will yield us a higher-order derivative Lagrangian, that is, a Lagrangian that contains the second derivative of $a$. Such a Lagrangian can be canonical quantized by the Ostrogradsky scheme (see, for example, \cite{Ostrogradsky}).

However, we shall adopt the method proposed by Vilenkin \cite{Vilenkin:1985md} and use a Lagrangian multiplier $\lambda$ to rewrite the action (\ref{eqn:oriaction}) under the flat FRW metric (\ref{eqn:FRW}), where the expression of $R$ (\ref{eqn:Ricci}) is inserted as a constraint
\begin{equation}\label{eqn:mulaction}
 S=\int dt \left\{a^3f(R,L_m)
 -\lambda\left(R-6(\frac{\ddot{a}}{a}+\frac{\dot{a}^2}{a^2})\right)\right\}
\end{equation}
The action (\ref{eqn:mulaction}) can be regarded as a dynamical system described by two independent variables, the scale factor $a$ and the scalar curvature $R$. Varying (\ref{eqn:mulaction}) with respect to $R$ gives the value of $\lambda$
\begin{equation}\label{eqn:lambda}
 \lambda=a^3f_R
\end{equation}
We substitute (\ref{eqn:lambda}) back into (\ref{eqn:mulaction}) and integrate by parts to remove the second derivative of $a$, then obtain
\begin{multline}\label{eqn:redaction}
 S=\int dt \left[ a^3(f-f_RR)-6a\dot{a}^2f_R \right. \\
 \left. -6a^2\dot{a}\dot{R}f_{RR}-6a^2\dot{a}^2f_{RL_m}\frac{\partial L_m}{\partial a} \right]
\end{multline}
where we have assumed, as reviewd in Sec.~\ref{sec:fgravity}, that the matter Lagrangian density $L_m$ depends on the metric $g_{\mu\nu}$ and not on its derivatives. The reduced Lagrangian is
\begin{multline}\label{eqn:redlag}
 L(a,\dot{a},R,\dot{R})= a^3(f-f_RR)-6a\dot{a}^2f_R \\
 -6a^2\dot{a}\dot{R}f_{RR}-6a^2\dot{a}^2f_{RL_m}\frac{\partial L_m}{\partial a}
\end{multline}
where
\begin{equation}
 f_{RR}=\frac{\partial^2 f(R,L_m)}{\partial R^2}, \quad
 f_{RL_m}=\frac{\partial^2 f(R,L_m)}{\partial L_m\partial R} 
\end{equation}
Thus we get the reduced Lagrangian in which the FRW scale factor $a$ and the Ricci scalar $R$ play the role of independent dynamical variables.

It is straight forward to verify that the reduced action (\ref{eqn:redaction}) is equivalent to the original $f(R,L_m)$ gravity action (\ref{eqn:oriaction}) under the flat FRW metric (\ref{eqn:FRW}).


The variation of (\ref{eqn:redaction}) with respect to the scale factor $a$ gives
\begin{equation}\label{eqn:redfieldeq}
\begin{split}
 \frac{\delta S}{\delta a}
 &= \frac{\partial L}{\partial a}-\frac{d}{dt}\frac{\partial L}{\partial \dot{a}} \\
 &= 3fa^2 - 3f_Ra^2R + 6f_R\dot{a}^2 + 12f_Ra\ddot{a} \\
 & \quad + 12f_{RR}a\dot{a}\dot{R} + 6f_{RR}a^2\ddot{R} + 6f_{RRR}a^2\dot{R}^2 \\
 & \quad + f_{L_m}\frac{\partial L_m}{\partial a}a^3 - f_{RL_m}\frac{\partial L_m}{\partial a}a^3R \\
 & \quad + 18f_{RL_m}\frac{\partial L_m}{\partial a}a\dot{a}^2 
         + 12f_{RL_m}\frac{\partial L_m}{\partial a}a^2\ddot{a} \\
 & \quad + 6f_{RL_m}\frac{\partial^2 L_m}{\partial a^2}a^2\dot{a}^2
         + 12f_{RL_mR}\frac{\partial L_m}{\partial a}a^2\dot{a}\dot{R} \\
 & \quad + 6f_{RL_mL_m}\left(\frac{\partial L_m}{\partial a}\right)^2a^2\dot{a}^2
\end{split}
\end{equation}
Using the following mathematical identities
\begin{equation}
 \square f_R=\frac{1}{\sqrt{-g}}\partial_\mu(\sqrt{-g}g^{\mu\nu}\partial_\nu f_R)
\end{equation}
and
\begin{equation}
 \nabla_\mu\nabla_\nu f_R
 = \partial_\mu\partial_\nu f_R -\Gamma_{\mu\nu}^\lambda\partial_\lambda f_R
\end{equation}
one can directly verify that (\ref{eqn:redfieldeq}) implies the space-space component of the field equation (\ref{eqn:orifieldeq}) under the flat FRW metric (\ref{eqn:FRW}).

The variation of (\ref{eqn:redaction}) with respect to the Ricci scalar $R$ gives
\begin{equation}
\begin{split}
 \frac{\delta S}{\delta R}
 &= \frac{\partial L}{\partial R}-\frac{d}{dt}\frac{\partial L}{\partial \dot{R}} \\
 &= -a^3f_{RR}\left[R-6\left(\frac{\ddot{a}}{a}+\frac{\dot{a}^2}{a^2}\right)\right]
\end{split}
\end{equation}
which implies (\ref{eqn:Ricci}), the expression for Ricci scalar under the flat FRW metric (\ref{eqn:FRW}).

Thus, we explicitly verify that the reduced action (\ref{eqn:redaction}) is equivalent to the original $f(R,L_m)$ gravity model in such a way that the variations with respect to its dynamical variables $a$ and $R$ give the correct equations of motion as that of the $f(R,L_m)$ action (\ref{eqn:oriaction}) under the flat FRW metric (\ref{eqn:FRW}).


\section{Wheeler-DeWitt equation}\label{sec:wdequation}

Now we can canonical quantize the reduced $f(R,L_m)$ action (\ref{eqn:redaction}) in minisuperspace, and give a detailed derivation of the corresponding Wheeler-DeWitt equation
\begin{equation}
 \hat{H}\Psi=0
\end{equation}
which plays the role of Schr\"{o}dinger equation for the wave function of the universe in canonical quantum gravity.

We first introduce the following two new variables to diagonalize the derivative part of the reduced Lagrangian (\ref{eqn:redlag})
\begin{equation}\label{eqn:newvar}
 u \equiv a\sqrt{f_R}, \quad v \equiv \frac{1}{2}\ln f_R
\end{equation}
in terms of which the reduced Lagrangian (\ref{eqn:redlag}) reads
\begin{multline}
 L(u,\dot{u},v,\dot{v}) \\
 =\left(\frac{u}{\sqrt{f_R}}\right)^3
 \left[-6f_R\left(\frac{\dot{u}^2}{u^2}-\dot{v}^2\right)+(f-f_RR)\right]
\end{multline}
where $R$ should be expressed in terms of $v$ through (\ref{eqn:newvar}) once the exact form of $f(R,L_m)$ is given. The corresponding conjugate momenta are
\begin{subequations}\label{eqn:uvcamomenta}
\begin{align}
 P_u &= \frac{\partial L}{\partial \dot{u}} = -12\frac{u\dot{u}}{\sqrt{f_R}} \\
 P_v &= \frac{\partial L}{\partial \dot{v}} = 12\frac{u^3}{\sqrt{f_R}}\dot{v}
\end{align}
\end{subequations}
and the resulting Hamiltonian is
\begin{equation}\label{eqn:uvHam}
\begin{split}
 H &= \dot{u}P_u+\dot{v}P_v-L \\
 &= -\frac{1}{24}\frac{\sqrt{f_R}}{u}P_u^2 + \frac{1}{24}\frac{\sqrt{f_R}}{u^3}P_v^2 
 -\left(\frac{u}{\sqrt{f_R}}\right)^3(f-f_RR)
\end{split}
\end{equation}

The DeWitt metric of the minisuperspace spanned by $A,B=\{u,v\}$ is
\begin{equation}\label{eqn:demetric}
 G_{AB} = \left( \begin{array}{cc}
 \displaystyle -12\frac{u}{\sqrt{f_R}} & 0 \\
 0 & \displaystyle 12\frac{u^3}{\sqrt{f_R}}
 \end{array}\right)
\end{equation}
by which the Hamiltonian (\ref{eqn:uvHam}) can be written as
\begin{equation}\label{eqn:DeHam}
 H=\frac{1}{2}G^{AB}P_AP_B + V(u,v)
\end{equation}
where $G^{AB}$ is the inverse DeWitt metric and
\begin{equation}
 V(u,v)= -\left(\frac{u}{\sqrt{f_R}}\right)^3(f-f_RR)
\end{equation}
Since the Hamiltonian (\ref{eqn:DeHam}) is classically constrained to vanish because of the time-time component of the field equation (\ref{eqn:orifieldeq}), the corresponding operator $\hat{H}$ in the quantum theory must annihilate the wave function.


Making the following replacement in the Hamiltonian (\ref{eqn:DeHam})
\begin{equation}
 P_u \rightarrow -i\frac{\partial}{\partial u}, \quad
 P_v \rightarrow -i\frac{\partial}{\partial v}
\end{equation}
with the covariant Laplace-Beltrami ordering \cite{Kiefer:2008sw}
\begin{equation}
 G^{AB}P_AP_B \rightarrow -\nabla_{LB}^2 = -\frac{1}{\sqrt{-G}}\partial_A(\sqrt{-G}G^{AB}\partial_B)
\end{equation}
we obtain the Hamiltonian operator
\begin{multline}
 \hat{H} = \frac{1}{24}\frac{\sqrt{f_R}}{u^2}\frac{\partial}{\partial u}u\frac{\partial}{\partial u}
 -\frac{1}{24}\frac{\sqrt{f_R}}{u^3}\frac{\partial^2}{\partial v^2} \\
 -\left(\frac{u}{\sqrt{f_R}}\right)^3(f-f_RR)
\end{multline}
Imposing $\hat{H}\Psi=0$ gives the Wheeler-DeWitt equation
\begin{multline}\label{eqn:wdequation}
 \left[\frac{1}{24}\frac{\sqrt{f_R}}{u^2}\frac{\partial}{\partial u}u\frac{\partial}{\partial u}
 -\frac{1}{24}\frac{\sqrt{f_R}}{u^3}\frac{\partial^2}{\partial v^2}\right. \\
 \left.-\left(\frac{u}{\sqrt{f_R}}\right)^3(f-f_RR)\right]\Psi(u,v)=0
\end{multline}
where, as stated before, the Ricci scalar $R$ should be expressed in terms of $v$ through (\ref{eqn:newvar}) once the exact form of $f(R,L_m)$ is given.

The Wheeler-DeWitt equation (\ref{eqn:wdequation}), which can be regarded as an analogue of the Schr\"odinger equation in quantum $f(R,L_m)$ gravity, is a partial differential equation of the wave function $\Psi(u,v)$ in minisuperspace. As a result of quantized diffeomorphism-invariant theory, equation (\ref{eqn:wdequation}) is defined on the geometrical and matter degrees of freedom of the model, and not on spacetime points. It contains terms that diverge as the big bang (or the big crunch) is approached, $u=a\sqrt{f_R}\rightarrow 0$ as expected.


\section{Conclusion}\label{sec:conclusion}

In this paper we have derived the Wheeler-DeWitt equation of $f(R,L_m)$ gravity in flat FRW universe, where the three-geometry is homogeneous and isotropic. Since we are not dealing with inhomogeneous fluctuations of the metric, the infinite number of gravitational degrees of freedom are reduced to just two variables, the FRW scale factor $a$ and the Ricci scalar $R$. The reduced action of $f(R,L_m)$ gravity in the truncated minisuperspace can be given via a Lagrangian multiplier and then integration by parts. The equivalence of the reduced action and the original action are explicitly verified. Using the canonical quantization procedure, the Wheeler-DeWitt equation of $f(R,L_m)$ gravity is obtained. The Wheeler-DeWitt equation plays the role of Schr\"odinger equation for the wave function of the universe, and is at the heart of every canonical approach of quantum cosmology. Thus our work can serve as the first step of the study of quantum $f(R,L_m)$ cosmology. \medskip


\section{Acknowledgments}

I would like to thank David Atkatz for emailing me his paper on quantum cosmology \cite{quancos}, and Tomi Koivisto for email correspondence. This work was supported by the National Natural Science Foundation of China (NSFC) under No.\ 11247309, the Excellent Young Teachers Program of Universities in Shanghai (No.\ shu11028), and the Innovating Funds of Shanghai university.


\end{document}